\documentclass[12pt]{article}
\usepackage{amssymb}
\def\be{\begin{equation}}
\def\ee{\end{equation}}
\def\ba{\begin{array}}
\def\ea{\end{array}}
\def\bea{\begin{eqnarray}}
\def\eea{\end{eqnarray}}
\def\p{\partial}

\def\half{{\textstyle{1\over2}}}

\def\pl#1{{\sl Phys.~Lett.~\bf B#1}}
\def\pr#1{{\sl Phys.~Rev.~\bf D#1}}
\def\prl#1{{\sl Phys.~Rev. Lett.~\bf #1}}

\def\cqg#1{{\sl Class.~Quant.~Grav.~\bf #1}}

\catcode`\@=11
\def\@citex[#1]#2{%
\if@filesw \immediate \write \@auxout {\string \citation {#2}}\fi
\@tempcntb\m@ne \let\@h@ld\relax \def\@citea{}%
\@cite{%
  \@for \@citeb:=#2\do {%
    \@ifundefined {b@\@citeb}%
      {\@h@ld\@citea\@tempcntb\m@ne{\bf ?}%
      \@warning {Citation `\@citeb ' on page \thepage \space undefined}}%
      {\@tempcnta\@tempcntb \advance\@tempcnta\@ne%
      \@tempcntb\number\csname b@\@citeb \endcsname \relax%
      \ifnum\@tempcnta=\@tempcntb %Number follows previous--hold on to it
        \ifx\@h@ld\relax%
          \edef \@h@ld{\@citea\csname b@\@citeb\endcsname}%
        \else%
          \edef\@h@ld{\ifmmode{-}\else--\fi\csname b@\@citeb\endcsname}%
        \fi%
      \else%   %  non-successor--dump what's held and do this one
        \@h@ld\@citea\csname b@\@citeb \endcsname%
        \let\@h@ld\relax%
      \fi}%
    \def\@citea{,\penalty\@highpenalty\,}%
  }\@h@ld
}{#1}}

\def\@citeb#1#2{{[#1]\if@tempswa , #2\fi}}
\def\@citeu#1#2{{$^{#1}$\if@tempswa , #2\fi }}
\def\@citep#1#2{{#1\if@tempswa , #2\fi}}

\def\bcites{         % cite with []'s
        \catcode`\@=11
        \let\@cite=\@citeb
        \catcode`\@=12
}

\def\upcites{         % cite with exponents
        \catcode`\@=11
        \let\@cite=\@citeu
        \catcode`\@=12
}

\def\plaincites{      % cite without brackets
        \catcode`\@=11
        \let\@cite=\@citep
        \catcode`\@=12
}

\newcount\hour
\newcount\minute
\newtoks\amorpm
\hour=\time\divide\hour by 60
\minute=\time{\multiply\hour by 60 \global\advance\minute by-\hour}
\edef\standardtime{{\ifnum\hour<12 \global\amorpm={am}%
        \else\global\amorpm={pm}\advance\hour by-12 \fi
        \ifnum\hour=0 \hour=12 \fi
        \number\hour:\ifnum\minute<10 0\fi\number\minute\the\amorpm}}
\edef\militarytime{\number\hour:\ifnum\minute<10 0\fi\number\minute}

\def\draftlabel#1{{\@bsphack\if@filesw {\let\thepage\relax
   \xdef\@gtempa{\write\@auxout{\string
      \newlabel{#1}{{\@currentlabel}{\thepage}}}}}\@gtempa
   \if@nobreak \ifvmode\nobreak\fi\fi\fi\@esphack}
        \gdef\@eqnlabel{#1}}
\def\@eqnlabel{}
\def\@vacuum{}
\def\marginnote#1{}
\def\draftmarginnote#1{\marginpar{\raggedright\scriptsize\tt#1}}
\def\draft{
        \pagestyle{plain}
        \overfullrule=2pt
        \oddsidemargin -.5truein
        \def\@oddhead{\sl \phantom{\today\quad\militarytime} \hfil
        \smash{\Large\sl DRAFT} \hfil \today\quad\militarytime}
        \let\@evenhead\@oddhead
        \let\label=\draftlabel
        \let\marginnote=\draftmarginnote
        \def\ps@empty{\let\@mkboth\@gobbletwo
        \def\@oddfoot{\hfil \smash{\Large\sl DRAFT} \hfil}
        \let\@evenfoot\@oddhead}
        \def\@eqnnum{(\theequation)\rlap{\kern\marginparsep\tt\@eqnlabel}%
        \global\let\@eqnlabel\@vacuum}  }
\begin{document}

\hfill UTHET-04-0201

%\hfill {\tt hep-th/yymmxxx} 
\vspace{-0.2cm}

\begin{center}
\Large
{\bf On quasi-normal modes and the AdS$_5$/CFT$_4$ correspondence}
\normalsize

\vspace{0.8cm}
{\bf George Siopsis}
\footnote{
Research supported by the US Department of Energy under grant
DE-FG05-91ER40627.}
\\ Department of Physics
and Astronomy, \\
The University of Tennessee, Knoxville, \\
TN 37996 - 1200, USA. \\
{\tt email: siopsis@tennessee.edu}

\end{center}

\vspace{0.8cm}
\large
\centerline{\bf Abstract}
\normalsize
\vspace{.5cm}

We discuss the quasi-normal modes of massive scalar perturbations of black holes in AdS$_5$ in conjunction with the AdS/CFT correspondence.
On the gravity side, we solve the wave equation and obtain an expression for the asymptotic form of quasi-normal frequencies.
We then show that these expressions agree with those obtained from a CFT defined
on $\mathbb{R} \times S^3$ in a certain scaling limit, by identifying Euclidean
time with one of the periodic coordinates.
This generalizes known exact results in three dimensions (BTZ black hole).
As a by-product, we derive the standard energy quantization condition in AdS
by a simple monodromy argument in complexified AdS space. This argument relies on
an unphysical singularity.

\newpage
\section{Introduction}

Despite a considerable amount of work on
quasi-normal modes of black holes in asymptotically AdS space-times~\cite{bibq1,bibq2,bibq3,bibq4,bibq5,bibq6,bibq7,bibq8,bibq9,bibq10,bibq11,bibq12,bibq13,bibw1,bibw2,bibw6,bibw3,bibw4,bibw5},
their relation to the AdS/CFT correspondence is not well understood.
They are derived as complex eigenvalues corresponding to a wave equation which
is solved subject to the
conditions that the flux be in-going at the horizon and the wave-function vanish
at the boundary of AdS space.
The wave equation reduces to a hypergeometric equation in three dimensions (AdS$_3$)
and can therefore be solved exactly~\cite{bibq7,bibq13}.
In this case, the quasi-normal frequencies are the poles of the retarded
Green function of the corresponding perturbations in the dual CFT~\cite{bibq13}.
In higher dimensions, an analytic solution is not readily available, as the
wave equation develops unphysical singularities.
It has been possible to obtain numerical values for quasi-normal frequencies~\cite{bibq2,bibq14,bibr1,bibr2}. However, analytical expressions are needed in
order to understand their relevance to the AdS/CFT correspondence.

We recently developed an analytic method of calculating
quasi-normal modes of
large AdS black holes (high Hawking temperature $T_H$) in five dimensions.
Our method was based on a perturbative expansion of the wave equation,
which reduced to the Heun equation~\cite{bibq15}. We applied the method to the calculation of low-lying
frequencies~\cite{bibus} as well as in the asymptotic regime~\cite{bibus2} of
massless scalars.
Our results were in agreement with numerical results~\cite{bibq2,bibq14}.
We then extended the discussion of~\cite{bibus,bibus2} to the case of massive scalar modes~\cite{bibego}
and derived a systematic analytic expansion of the frequencies in powers of $1/m$,
where $m$ is the mass of the mode.
To lowest order, we obtain the asymptotic expression
\be\label{eqA} \frac{\omega_n}{T_H} \sim 2\pi (\pm 1 -i) (n+\half \Delta_+ - {\textstyle{\frac{3}{2}}})
\ \ , \ \ n = 1,2,\dots \ee
where $\Delta_\pm = 2\pm \sqrt{4+ m^2R^2}$ and $R$ is the AdS radius,
confirming earlier numerical results~\cite{bibns}.

The large real part of the quasi-normal frequencies~(\ref{eqA}) challenges our understanding
of the AdS/CFT correspondence in five dimensions.
The quasi-normal frequencies determine the poles of the retarded correlation
functions of dual operators in finite-temperature $\mathcal{N} = 4~SU(N)$ SYM theory in the large-$N$, large 't Hooft coupling limit~\cite{bibns}.
They have also been recently studied in~\cite{bibother} where they were shown to arise in complexified geodesics (see also~\cite{bibother2}).
Here we show that the quasi-normal frequencies~(\ref{eqA}) are obtained as the
poles of a CFT on $\mathbb{R}\times S^3$ in a certain scaling limit where
Euclidean time is identified with one of the periodic coordinates of $S^3$.
This generalizes exact known results in three dimensions (BTZ black hole~\cite{bibq13}) where the CFT is defined on the cylinder $\mathbb{R}\times S^1$.

We also discuss the geometric origin of our results in terms of the
Riemann surfaces which solve the corresponding wave equations~\cite{bib26}.
In three dimensions, the equivalence of the two Riemann surfaces (BTZ and AdS$_3$) follows directly from the equivalence of the corresponding metrics.
In five dimensions, the black hole and AdS metrics are not equivalent.
We shall show that the corresponding Riemann $w$-surfaces
are equivalent in certain limits we discuss, even though the
corresponding metrics are not.

Our work is organized as follows. In section~\ref{sec2}, we review the three-dimensional
case where exact results are obtained.
In section~\ref{sec3}, we discuss the five-dimensional case. We obtain the
asymptotic form of the quasi-normal frequencies for a massive scalar analytically,
and show that they agree with the poles of a propagator obtained in the dual
CFT which lives on the boundary $\mathbb{R}\times S^3$ in a certain scaling limit.
We also obtain the standard energy quantization condition in AdS$_5$ by a
simple monodromy argument which, however, relies on an unphysical singularity.
Our conclusions are summarized in section~\ref{sec4}.

\section{Three dimensions}\label{sec2}

Here we review known exact results in three dimensions (BTZ black hole~\cite{bibq7,bibq13}) for later comparison with the five-dimensional case.
The metric for a non-rotating BTZ black hole reads
\be\label{eqmbtz}
ds^2 = - \left( r^2 -r_h^2 \right) dt^2 +\frac{dr^2}{ \left( r^2 - r_h^2 \right) }+ r^2 dx^2
\ee
where $r_h$ is the radius of the horizon (we
set the AdS radius $R=1$).
The wave equation for a massive scalar of mass $m$ is
\be
\frac{1}{r}\p_r \left( r^3 \left( 1- \frac{r_h^2}{r^2}\right) \p_r \Phi\right) -\frac{1}{r^2 - r_h^2 }\p_t^2 \Phi + \frac{1}{r^2}\p_x^2 \Phi = m^2\Phi
\ee
%One normally solves this equation in the physical interval $r\in [r_h,\infty)$.
%Instead, we shall solve it in the interval $0\le r\le r_h$ (inside the horizon).
The solution may be written as
\be
\Phi = e^{-i(\omega t-px) }\Psi (y) ,\ \ \ \ \ y = \frac{r^2}{r_h^2}
\ee
where $\Psi$ satisfies
\be\label{eq5a}
\left( y (1-y) \Psi' \right)'
+ \left( \frac{\hat\omega^2}{1- y} +\frac{\hat p^2}{ y} \right)\Psi + \frac{m^2}{4} \Psi =0
\ee
%in the interval $0<y<1$, 
and we have introduced the dimensionless variables
\be
\hat\omega = \frac{\omega}{2r_h} = \frac{\omega}{4\pi T_H},\ \ \ \hat p = \frac{p}{2r_h} = \frac{p}{4\pi T_H}
\ee
where $T_H = r_h/(2\pi)$ is the Hawking temperature.
Two independent solutions are obtained by examining the behavior near the horizon
($y\to 1$),
\be\label{eq11} \Psi_\pm \sim (1-y)^{\pm i\hat\omega}\ee
where $\Psi_+$ is outgoing and $\Psi_-$ is in-going. A different set of linearly
independent solutions is obtained by studying the behavior at the black hole
singularity ($y\to 0$).
We obtain
\be \Psi\sim y^{\pm i\hat p}
\ee
For quasi-normal modes, we demand that $\Psi$ be purely in-going at the horizon
($\Psi \sim \Psi_-$ as $y\to 1$). By writing
\be \Psi (y) = y^{i\hat p} (1-y)^{-i\hat\omega} F(y)\ee
we deduce
\be\label{eqhy0} y(1-y) F'' + \{ 1+ 2i\hat p -(2-2i(\hat\omega -\hat p)y \}\, F'
+ \left\{ (\hat\omega - \hat p)(\hat\omega -\hat p +i) + \frac{m^2}{4} \right\} \, F =0 \ee
The solution which is regular at the horizon ($y\to 1$) is the
Hypergeometric function
\be\label{eqhy} F (y) = F (a_+,a_-;c; 1-y)
\ \ , \ \ a_\pm = \half\Delta_\pm -i(\hat\omega - \hat p) \ , \ c= 1-
2i\hat\omega\ee
where
\be \Delta_\pm = 1\pm \sqrt{1+m^2} \ee
%In general, near the horizon ($y\to 1$), this solution is a mixture of in-going and outgoing waves.
As $y\to\infty$, this function behaves as
\be F(y) \sim \mathcal{A} y^{-a_+} + \mathcal{B} y^{-a_-}
\ee
where
\be \mathcal{A} = \frac{\Gamma (c)\Gamma (a_--a_+)}{\Gamma (a_-)\Gamma (c-a_+)}\ \ , \ \ \mathcal{B} = \frac{\Gamma (c)\Gamma (a_+-a_-)}{\Gamma (a_+)\Gamma (c-a_-)}\ee
For the desired behavior at infinity ($\Psi\sim y^{-\Delta_+/2}$), we ought to
set
\be \mathcal{B} = 0\ee
%It also blows up at infinity. To obtain the desired behavior for a quasi-normal
%mode at $y=1$ and $y\to\infty$, we demand that $F(y)$ be a Polynomial.
This condition implies
\be\label{eqwme2} \hat\omega = \pm \hat p  -i(n +\half\Delta_+ -1)\quad,\quad n=1,2,\dots\ee
a discrete set of complex frequencies with negative imaginary part, as expected~\cite{bibq2}. Notice that we obtained two sets of frequencies, with opposite real parts.
%Then  eq.~(\ref{eqhy}) implies that
%\be F(y) = {}_2F_1 (1-n,-n; 1\pm 2i\hat p; y)\ee
%which is a Polynomial of order $n-1$.
%It is therefore a constant at $y=1$, as desired and behaves as $F(y)\sim y^{n-1}
%\sim y^{i(\hat\omega \mp \hat p)-1}$ as $y\to\infty$. Therefore,
%$\Psi\sim y^{-1}$ as $y\to\infty$, as expected.

It is interesting to note that the above quasi-normal frequencies may also be deduced from a simple monodromy
argument.
It is based on a nice geometrical interpretation of the solution to the wave equation~\cite{bib26}.
First, let us bring the wave equation~(\ref{eq5a}) into a Schr\"odinger-like form
by introducing the ``tortoise'' coordinate
\be z(y) = \int \frac{dy}{\sqrt y (1-y)} = \ln \frac{\sqrt y - 1}{\sqrt y + 1} -i\pi \ee
where the constant was fixed so that the black hole singularity $y=0$ is mapped onto $z=0$.
The wavefunction
\be \varphi (z) = y^{1/4} \Psi (y) \ee
satisfies the wave equation
\be\label{eq19} - \frac{d^2\varphi}{dz^2} + V[y(z)] \varphi = \hat\omega^2 \varphi \ \ , \ \ V(y) = (y-1) \left\{ \frac{\hat p^2 + \frac{1}{16}}{y} + \frac{3}{16} + \frac{m^2}{4} \right\} \ee
Let $\varphi_\pm$ be two linearly independent solutions of~(\ref{eq19}).
Their ratio defines a map from the complex $z$-plane to the complex $w$-plane,
\be\label{eqmap} z \mapsto w(z) \equiv \frac{\varphi_+ (z)}{\varphi_- (z)} \ee
In terms of $w$, the wave equation takes a simple form,
\be \frac{d^2\varphi}{dw^2} = 0 \ee
Thus, solving the wave equation amounts to finding the Riemann surface which is the image of the complex $z$-plane under the map~(\ref{eqmap}).
The $w$-surface inherits the singularities of the wave equation and is uniquely
determined by the monodromies around them.

The quasi-normal modes may be obtained by simply looking at the monodromies around the singular points of the $w$-surface~\cite{bib26}.
They coincide with the monodromies computed through the hypergeometric equation~(\ref{eqhy0}),
the latter being equivalent to eq.~(\ref{eq19}).
We may choose $\varphi_\pm = y^{1/4} \Psi_\pm$, and then
\be w = \frac{\Psi_+}{\Psi_-} \ee
where $\Psi_\pm$ are as in~(\ref{eq11}).
Let $\mathcal{M}(y_0)$ be the monodromy around the singular point $y=y_0$ computed along a small circle centered at $y=y_0$ running counterclockwise.
For $y=1, 0 ,\infty$, we obtain, respectively,
\be\label{eqmono3}\mathcal{M} (1) = \left( \ba{cc} e^{2\pi\hat\omega} & \\ & e^{-2\pi\hat\omega} \ea \right) \ \ , \ \ \mathcal{M} (0) = \left( \ba{cc} e^{2\pi \hat p} & \\ & e^{-2\pi \hat p} \ea \right) \ \ , \ \ \mathcal{M} (\infty)
= \left( \ba{cc} e^{i\pi \Delta_+} & \\ & e^{i\pi \Delta_-}\ea \right) \ee
For quasi-normal modes, we must have~\cite{bib26}
\be\label{eqmo1} e^{-2\pi\hat\omega} e^{\pm 2\pi \hat p} = e^{i\pi \Delta_+} \ee
which leads to the same set of quasi-normal frequencies as before
(eq.~(\ref{eqwme2})).
The above result~(\ref{eqmo1}) may also be obtained directly by working in the complex $y$-plane
and deforming the contours around $y=0$ and $y=1$ to the contour at infinity.
%Notice that the monodromy argument is much simpler here than in the case of
%an asymptotically flat space-time~\cite{bibmo2}.
%This is because of a simpler boundary condition at infinity ($y\to \infty$).

Turning to the AdS/CFT correspondence, the flux at the boundary ($y\to\infty$)
is related to the retarded propagator of the corresponding CFT living on
the boundary. A standard calculation yields
\be \tilde G_R (\omega, p) \sim \lim_{y\to\infty} \frac{F'(y)}{F(y)}\ee
Explicitly,
\bea\label{eq20}
\tilde G_R (\omega, p) &\sim& \frac{\mathcal{A}}{\mathcal{B}}\nonumber\\
&\sim& \frac{\Gamma (\half \Delta_+ - i(\hat\omega -\hat p))\Gamma (\half \Delta_+ - i(\hat\omega +\hat p))}{\Gamma (1-\half \Delta_+ - i(\hat\omega -\hat p))\Gamma (1-\half \Delta_+ - i(\hat\omega +\hat p))}\nonumber\\
&\sim& |\Gamma (\half \Delta_+ - i(\hat\omega -\hat p))\Gamma (\half \Delta_+ - i(\hat\omega +\hat p))|^2 \nonumber\\
& &\times \sin\pi(\half \Delta_+ - i(\hat\omega -\hat p))
\sin\pi(\half \Delta_+ - i(\hat\omega +\hat p))\eea
Plainly, the quasi-normal modes (zeroes of $\mathcal{B}$) are poles of the
retarded propagator (since $G_R\sim 1/\mathcal{B}$).

We observe that the above quantization
condition~(\ref{eqwme2}) may be derived from AdS$_3$, even though the latter
is associated with zero temperature. To see this, consider Euclidean AdS$_3$
whose metric may be written as
\be\label{eqmads} ds^2 = \cosh^2\rho d\tau^2 + d\rho^2 +\sinh^2\rho d\phi^2\ee
For finite temperature, we identify the periodic coordinate $\phi$ with Euclidean time,
\be t_E = \frac{\phi}{2\pi T}\ee
so that $t_E$ has period $1/T$, where $T$ is to be identified with temperature.
The other coordinate $\tau$ is then a spatial coordinate. The boundary on which
the corresponding CFT lives is the cylinder
$\mathbb{R}\times S^1$.

Upon a change of coordinates,
\be y = \cosh^2\rho \ \ , \ \ x = \frac{\tau}{2\pi T}\ee
the metric~(\ref{eqmads}) becomes
\be ds^2 = (2\pi T)^2 (y-1) dt_E^2 + \frac{dy^2}{4y(y-1)} + (2\pi T)^2 ydx^2\ee
which is identical to the Wick-rotated BTZ black hole metric~(\ref{eqmbtz}) with $y = r^2/r_h^2$, $r_h = 2\pi T$.

To compare with the corresponding CFT, it is advantageous to write the propagator in coordinate space.
This is most easily done through the invariant distance in the embedding
\be \mathcal{P}(X, X') = -(X^0-X^{\prime 0})^2 + (X^1-X^{\prime 1})^2 + (X^2-X^{\prime 2})^2 + (X^3-X^{\prime 3})^2\ee
where
\be X^0 = \cosh\rho \cosh\tau\ \ , \ \ X^3 = \cosh\rho \sinh\tau\ee
\be X^1 = \sinh\rho \cos\phi \ \ , \ \ X^2 = \sinh\rho \sin\phi\ee
and similarly for $X'$. The (Euclidean) propagator on the boundary is found to be
\be G (\tau, \phi; \tau',\phi') \sim \lim_{\rho,\rho'\to\infty} \mathcal{P}^{-\Delta_+/2}\ee
In this limit,
\be\label{eqPcal} \mathcal{P} \sim \cosh(\tau - \tau') -\cos(\phi - \phi')\ee
therefore,
\be\label{eq30} G (\tau, \phi; \tau',\phi') \sim \frac{1}{(\cosh(\tau - \tau') -\cos(\phi - \phi'))^{\Delta_+/2}}\ee
It is a straightforward exercise to show that this Euclidean propagator is the fourier transform of
the expression for the retarded propagator $\tilde G_R$~(\ref{eq20}) derived above,
after a Wick rotation back to Minkowski space.

The above result~(\ref{eq30}) may also be obtained by a direct
CFT calculation (without reference to the corresponding AdS).
To this end, observe that the two-point function of a massless scalar field $\Phi$ on the cylinder $\mathbb{R}\times S^1$ is
\be G_0(\tau, \phi; \tau',\phi') \equiv \langle T(\Phi (\tau, \phi) \Phi (\tau',\phi')) \rangle
\sim
T \sum_{j=-\infty}^\infty \int \left. \frac{dk}{2\pi}\ e^{-ik\cdot x}
\ \frac{i}{k^2}\right|_{k^0 = 2\pi jT}\ee
After integrating over $k$, summing over $j$ and subtracting an irrelevant (infinite) constant, we obtain
\be G_0(\tau, \phi; \tau',\phi') \sim \ln\mathcal{P}\ee
where $\mathcal{P}$ is given by eq.~(\ref{eqPcal}). For a scalar operator $\mathcal{O}$
of dimension $\Delta$, the two point function then reads
\be G(\tau, \phi; \tau',\phi') \equiv \langle T(\mathcal{O} (\tau, \phi) \mathcal{O} (\tau',\phi')) \rangle
\sim \frac{1}{\mathcal{P}^{\Delta/2}}\ee
in agreement with eq.~(\ref{eq30}).

\section{Five dimensions}\label{sec3}

Extending the above results to five dimensions is far from straightforward.
The wave equation in an AdS black hole background reduces to a Heun equation
which contains more singularities than a hypergeometric equation and is in general
unsolvable.
We discuss a perturbative solution to the wave equation for a massive scalar
of mass $m$ yielding an asymptotic
form of quasi-normal frequencies valid for large $m$~\cite{bibego}.
We then attempt to extend the three-dimensional results by considering pure
AdS$_5$ and its dual CFT$_4$ which lives on $\mathbb{R}\times S^3$.
We show that in a certain scaling limit, the quasi-normal frequencies are the
poles of the propagator in CFT$_4$.

\subsection{Quasi-normal modes}

We start by deriving the asymptotic form of quasi-normal modes analytically, following the discussion in~\cite{bibego}.
%We then obtain the first-order correction in the momentum of the mode.
%The $o(1/m)$ corrections discussed in~\cite{bibego} are not considered here.
The metric of a five-dimensional AdS black hole in the high-temperature
limit may be written as
\be
ds^2 = -r^2 h(r)\,  dt^2 + \frac{dr^2}{r^2 h(r)} +r^2 ds^2 (\mathbb{R}^3)
\ \ , \ \
h(r) = 1 - \frac{r_h^4}{r^4}\ee
We have set the AdS radius $R=1$ to simplify the notation.
The radius of the horizon $r_h$ is proportional to $M^{1/4}$, where $M$ is the
(large) mass of the black hole. The (high) Hawking temperature is given by
\be
T_H = \frac{r_h}{\pi}
\ee
The wave equation for a massive scalar of mass $m$ is
\be\label{eq5}
\frac{1}{r^3}\p_r (r^5\, h(r)\, \p_r \Phi) -\frac{1}{ r^2\, h(r) }\p_{t}^2\Phi + \frac{1}{r^2}\; \vec\nabla^2\Phi = m^2 \Phi
\ee
We are interested in solving this equation for
a wave which is ingoing at the horizon and vanishes at infinity. These boundary
conditions yield a discrete set of complex frequencies (quasi-normal modes).
The solution may be written as
\be 
\Phi = e^{-i(\omega t - \vec p\cdot \vec x)} \Psi (r)
\ee
Upon changing the coordinate $r$ to $y$,
\be
y = \frac{r^2}{r_h^2} 
\ee
the wave equation becomes
\be\label{eq24}
\left( y(y^2-1) \Psi' \right)' + \left(\hat\omega^2\ \frac{y^2}{y^2-1} - \hat p^2 - \frac{m^2}{4}\ y\right)\Psi = 0
\ee
where we have introduced the dimensionless variables
\be\label{eqhat}
\hat\omega = \frac{\omega}{2\pi T_H} \ \ , \ \ \ \ \
\hat p = \frac{|\vec p|}{2\pi T_H}
\ee
Before attempting to solve this equation, let us bring it to a Schr\"odinger-like form as in the three dimensional case.
We define the ``tortoise'' coordinate by
\be z(y) = \int \frac{\sqrt y dy}{y^2 - 1} = \frac{1}{2} \ln \frac{\sqrt y -1}{\sqrt y +1} + \frac{i}{2} \ln \frac{\sqrt y +i}{\sqrt y-i} \ee
The wavefunction
\be \varphi (z) = y^{3/4} \Psi(y) \ee
satisfies the wave equation
\be - \frac{d^2\varphi}{dz^2} + V[y(z)]\varphi = \hat\omega^2\varphi \ \ , \ \
V(y) = (y^2-1) \left\{ \frac{m^2 -\frac{15}{4}}{4y} + \frac{\hat p^2}{y^2} - \frac{9}{16y^3} \right\} \ee
We now have four singular points, $y=0, \pm1, \infty$,
rendering the wave equation unsolvable analytically.
We can solve it in the high-energy regime by finding an approximation which
is valid around the singular point $y=\infty$.
This will effectively eliminate the singularity at $y=0$ and
simplify the Riemann $w$-surface~(\ref{eqmap}).

Two independent solutions are obtained by examining the behavior near the
horizon ($y\to 1$),
%Near the horizon we obtain in- and out-going waves
\be\label{eqn1} \Psi_\pm \sim (y-1)^{\pm i\hat\omega/2}\ee
where $\Psi_+$ is outgoing and $\Psi_-$ is ingoing.
We ought to choose $\Psi_-$ for quasi-normal modes.

A different set of linearly independent solutions is obtained by studying the
behavior at large $r$
($y\to \infty$). We obtain
\be\label{eq27} \Psi\sim y^{-\Delta_\pm/2} \ \ , \ \ \Delta_\pm = 2\pm \sqrt{4+m^2}\ee
For quasi-normal modes, we are interested
in the solution which vanishes at the boundary,
hence it behaves as $\Psi\sim y^{-\Delta_+/2}$ as $y\to\infty$.
Combining this with the requirement that $\Psi = \Psi_-$ (eq.~(\ref{eqn1}))
leads to a discrete spectrum of quasi-normal frequencies.

By considering the other (unphysical) singularity at $y=-1$, we obtain
yet another set of linearly independent wave-functions behaving as
\be\label{eq27u} \Psi \sim (y+1)^{\pm \hat\omega /2} \ee
near $y=-1$. There is no restriction on the behavior of the wave-function at this singularity.
Nevertheless, it is advantageous to isolate the behavior
near $y=-1$.

Isolating the behavior at the two singularities $y=\pm 1$, we shall
write the wave-function as
\be\label{eq25}
\Psi (y) = (y-1)^{-i\hat\omega/2} (y+1)^{-\hat\omega/2} F(y)
\ee
The selection of the exponent at the $y=-1$ singularity is arbitrary and not
guided by a physical principle. It leads to a convenient
perturbative calculation of {\em half} of the modes.
Selecting the other exponent, $+\hat\omega/2$, similarly yields the other set
of modes without additional computational difficulties. The latter have the
same imaginary part as the former, but opposite real parts.

It is easily deduced from eqs.~(\ref{eq24}) and (\ref{eq25}) that
the function $F(y)$ satisfies the Heun equation
\be\label{w1}
y(y^2-1) F'' + P(y) F' + Q(y) F = 0 \ee
where $P$ and $Q$ are polynomials in $y$,
\be\label{w1a}
P(y) = \left( 3- (1+i)\ \hat\omega \right) y^2 + (1-i)\ \hat\omega \ y -1 \ \ ,\ \
Q(y) = \hat\omega\ \left( \frac{i\hat\omega}{2} - 1-i\right) y -\frac{m^2}{4}\ y + (1-i)\frac{\hat\omega}{2} - \hat p^2
\ee
The high-energy limit corresponds to large $y$.
We may therefore drop the constant terms in the two polynomials $P$ and $Q$.
This approximation will yield an asymptotic form of the quasi-normal frequencies
in the high-energy limit. Indeed, viewed as functions of $\hat\omega$,
each constant term is of lower order in $\hat\omega$ than the rest of the polynomial.
%We wish to solve this equation in a region in the complex $y$-plane
%containing $|y|\ge 1$, which includes the
%physical regime $r> r_h$.
%For large $\hat\omega$, the constant terms in the respective polynomial coefficients of $F'$ and $F$ in~(\ref{w1}) are small compared with the other terms, so they may be dropped.
Eq.~(\ref{w1}) may then be approximated by the hypergeometric equation
\be\label{w2x}
(y^2-1) F'' + \left\{ \left( 3- (1+i)\ \hat\omega \right) y + (1-i)\ \hat\omega \right\} F'
+ \left\{ \hat\omega\ \left( \frac{i\hat\omega}{2} - 1-i\right)
- \frac{m^2}{4} \right\} \; F =0 \label{w2}
\ee
in the high-energy limit of large frequencies $\hat\omega$.
As anticipated, eq.~(\ref{w2}) contains only three singularities ($y=\pm 1,\infty$).

%The singular points $(1,-1,\infty)$ correspond to the horizon, an unphysical
%point and the boundary of AdS, respectively.
Two linearly independent solutions of (\ref{w2x}) are
\be\label{eqsolinf} \mathcal{K}_\pm = (x+1)^{-a_\pm} F(a_\pm, c-a_\mp ; a_\pm -a_\mp +1; 1/(x+1))\ee
where
\be\label{eqxy} a_\pm = \frac{\Delta_\pm - (1+i) \hat\omega}{2}
\quad,\quad c = \frac{3}{2} - i \hat\omega \ \ , \ \ x = \frac{y-1}{2}\ee
We have $\mathcal{K}_\pm\sim x^{-a_\pm}$ as $x\to\infty$, and correspondingly,
$\Psi\sim x^{-\Delta_\pm/2}$ (from eq.~(\ref{eq25}) using~(\ref{eqxy})).
Therefore,
the desired solution is $\mathcal{K}_+$, since it leads to $\Psi\to 0$ as $x\to\infty$.
Near the horizon ($x\to 0$), it behaves as
\be \mathcal{K}_+ = \mathcal{A} + \mathcal{B} x^{1-c}
\ee
where
\be\label{eq21} \mathcal{A} = \frac{\Gamma(1-c)\Gamma(1-a_-+a_+)}{\Gamma(1-a_-)\Gamma(1-c+a_+)}
\ \ , \ \ \mathcal{B} = \frac{\Gamma(c-1)\Gamma(1+a_+-a_-)}{\Gamma(a_+)\Gamma(c-a_-)}\ee
For regularity at the horizon, we demand
\be \mathcal{B} = 0\ee
which leads to a quantization condition yielding the
quasi-normal modes.
The quasi-normal frequencies are obtained as~\cite{bibego}
\be\label{eqo} \hat\omega_n = (- 1- i)(n+\half\Delta_+-{\textstyle{\frac{3}{2}}}) + o(1/m) \ \ , \ \ n=1,2,\dots\ee
in agreement with numerical results~\cite{bibns}.
The other set of modes (same as~(\ref{eqo}) but with opposite (positive) real part) is obtained by choosing the factor $(y+1)^{+\hat\omega/2}$, instead, in the {\em ansatz}~(\ref{eq25}) before approximating.

To better understand the geometric origin of these modes, we may construct an approximation to the $w$-surface~(\ref{eqmap}).
As explained earlier~\cite{bib26},
we only need the monodromies around the singular points of the $w$-surface to
determine it.
These monodromies can be computed from the approximate wave equation~(\ref{w2}).
The monodromies at the singular points of~(\ref{w2}), $y=\pm 1,\infty$ are
easily found to be
%may be computed from
%the approximate wave equation~(\ref{w2}).
%We obtain
\be\label{eqmono5}\mathcal{M} (1) = \left( \ba{cc} e^{\pi\hat\omega} & \\ & e^{-\pi\hat\omega} \ea \right) \ , \ \mathcal{M} (-1) = \left( \ba{cc} e^{i\pi \hat\omega } & \\ & e^{-i\pi (\hat\omega -1)} \ea \right) \ , \ \mathcal{M} (\infty)
= \left( \ba{cc} e^{i\pi \Delta_+} & \\ & e^{i\pi \Delta_-}\ea \right) \ee
For quasi-normal modes, we must have ({\em cf.}~eq.~(\ref{eqmo1}))
\be\label{eqmo15} e^{-\pi\hat\omega} e^{-i\pi (\hat\omega -1)} = e^{i\pi \Delta_+} \ee
leading to the expression~(\ref{eqo}) of frequencies with negative real part.
This result may also be obtained by contour deformation in the complex $y$-plane.

In a manner similar to three dimensions ({\em cf.}~eq.~(\ref{eq20})), we obtain the retarded Green function on
the boundary,
\be \tilde G_R \sim \frac{\mathcal{A}}{\mathcal{B}}\ee
whose poles are evidently the quasi-normal frequencies~(\ref{eqo}). However,
unlike in three dimensions, this is only an approximate expression with
corrections being of $o(1/m)$.

\subsection{Pure AdS and its dual CFT}

In three dimensions, the quasi-normal modes~(\ref{eqwme2}) were seen to
coincide with the poles of the Green function obtained from a CFT living
on the cylinder $\mathbb{R}\times S^1$ which is the boundary of AdS$_3$.
This was easily seen by considering Euclidean AdS$_3$ and identifying the
time coordinate with a periodic coordinate in EAdS$_3$. To extend this
discussion to five dimensions, we consider EAdS$_5$ and its dual CFT living
on the boundary $\mathbb{R}\times S^3$. The poles of the propagator will be
seen to coincide with the asymptotic quasi-normal frequencies~(\ref{eqo}) in
a certain scaling limit.

EAdS$_5$ is defined as the hyperboloid
\be X_0^2 - X_1^2 - X_2^2 - X_3^2 - X_4^2 - X_5^2 = 1\ee
in the flat six-dimensional Minkowski space with metric
\be ds^2 = -dX_0^2 + dX_1^2 + dX_2^2 + dX_3^2 + dX_4^2 + dX_5^2\ee
Use parameters
\bea X_0 &= \cosh\rho \cosh \tau \ \ , \ \ & X_5 = \cosh\rho\sinh\tau \ \ , \nonumber\\
X_1 &= \sinh\rho \cos\theta \cos \phi_1 \ \ , \ \ & X_2 = \sinh\rho \cos\theta \sin \phi_1 \ \ , \nonumber\\
X_3 &= \sinh\rho \sin\theta \cos \phi_2 \ \ , \ \ & X_4 = \sinh\rho \sin\theta \sin \phi_2
\eea
The metric on EAdS$_5$ then reads
\be ds^2 = \cosh^2\rho d\tau^2 + d\rho^2 + \sinh^2\rho \Big( \cos^2\theta d\phi_1^2 + d\theta^2 + \sin^2\theta d\phi_2^2
\Big)\ee
Switch variables to
\be u = \tanh^2\rho\ee
The metric becomes
\be\label{eq57} ds^2 = \frac{d\tau^2}{1-u} + \frac{du^2}{u(1-u)^2} + \frac{u}{1-u} \Big( \cos^2\theta d\phi_1^2 + d\theta^2 + \sin^2\theta d\phi_2^2
\Big)\ee
where $0\le u\le 1$, the boundary being at $u=1$.
The wave equation for a massive scalar of mass $m$ is
\be\label{eq60} u(1-u)\ \frac{\partial^2\Psi}{\partial u^2} + (2-u) \ \frac{\partial\Psi}{\partial u}
+ \frac{1}{4}\ \frac{\partial^2\Psi}{\partial\tau^2}
- \frac{\hat L^2}{4u}\ \Psi = \frac{m^2}{1-u}\ \Psi
\ee
where
\be\label{eqL2} \hat L^2 = -\frac{1}{\cos\theta\sin\theta} \ \frac{\partial}{\partial\theta} \left(
\cos\theta\sin\theta \frac{\partial}{\partial\theta} \right) - \frac{1}{\cos^2\theta}\ \frac{\partial^2}{\partial \phi_1^2} - \frac{1}{\sin^2\theta}\ \frac{\partial^2}{\partial \phi_2^2}\ee
has eigenfunctions
\be\label{eqL2e} Y_{Lj_1 j_2} (\theta,\phi_1,\phi_2) = e^{i(j_1\phi_1+j_2\phi_2)} \cos^{j_1}\theta
\sin^{j_2}\theta\ F(1+b_+, b_-; 1+j_1; \cos^2\theta)\ \ , \ \
b_\pm = \half(j_1 + j_2\pm L)\ee
and corresponding eigenvalues $L(L+2)$.

The solution to the wave equation is
\be \Psi = e^{iE\tau} u^{L/2} (1-u)^{\Delta_+/2} \ F(\alpha_+,\alpha_-;L+2;u)\ Y_{Lj_1 j_2} (\theta, \phi_1,\phi_2) \ \ , \ \ \alpha_\pm = \half( L + \Delta_+ \pm i E )\ee
The energy quantization condition is obtained by demanding $F$ be a polynomial.
Then $\alpha_\pm = -n+1$ ($n = 1,2,\dots$), leading to
\be\label{eqE} \pm iE = L+\Delta_++2n-2 \ \ , \ \ n = 1,2,\dots\ee
It is interesting to note that this quantization rule may also be obtained by a monodromy argument. Notice that
the wave equation~(\ref{eq60}) has singularities at $u = 0, 1, \infty$, the third one being unphysical, since $0\le u \le 1$. From the wave equation~(\ref{eq60})
we may find the behavior of the wavefunction near the singularities.
We therefore deduce the monodromies
\be\label{eqmoA}\mathcal{M} (0) = \left( \ba{cc} e^{i\pi (L+2)} & \\ & e^{- i\pi L} \ea \right) \ , \ \mathcal{M} (1) = 
\left( \ba{cc} e^{i\pi \Delta_+} & \\ & e^{i\pi \Delta_-}\ea \right)  \ , \ \mathcal{M} (\infty) = \left( \ba{cc} e^{-\pi E} & \\ & e^{\pi E} \ea \right) \ee
Demanding correct behavior in the interior ($u=0$) and at the boundary of AdS, we must have ({\em cf.}~eq.~(\ref{eqmo15}))
\be\label{eqmo1A} e^{\pm\pi E} e^{-i\pi L} = e^{i\pi \Delta_+} \ee
leading to the quantization condition~(\ref{eqE}).

Approaching the AdS boundary, $u\to 1$, we obtain $\Psi\sim (1-u)^{\Delta_+/2} \Phi$,
where
\be \Phi = e^{iE\tau} e^{i(j_1\phi_1+j_2\phi_2)} \cos^{j_1}\theta
\sin^{j_2}\theta\ F(1+b_+, b_-; 1+j_1; \cos^2\theta)\ee
is an eigenfunction of the Laplacian on $\mathbb{R}\times S^3$,
\be \Delta_{\mathbb{R}\times S^3} = \frac{\partial^2}{\partial \tau^2} - \hat L^2\ee
with eigenvalue $-E^2 - L(L+2)$.
They form a representation of the conformal group $SO(5,1)$. The quantum numbers $E$ and $L$ correspond to the subgroups $SO(1,1)$ (generated by $\partial/\partial\tau$) and $SO(4)$, the product $SO(1,1)\times SO(4)$ being a maximal
subgroup of the conformal group.

Turning to the CFT, we note that the propagator on the boundary $\mathbb{R}\times S^3$
is easily expressed in terms of the invariant
distance in the embedding of AdS$_5$,
\bea\label{eq68} \mathcal{P} &=& \lim_{\rho,\rho'\to\infty} (X-X')^A (X-X')_A\nonumber\\
&\sim& \cosh(\tau - \tau') - \cos\theta\cos\theta' \cos (\phi_1-\phi_1')
- \sin\theta\sin\theta' \cos (\phi_2-\phi_2')\eea
as
\be\label{eq69} G(x,x') \sim \frac{1}{\mathcal{P}^{\Delta_+/2}}\ee
where $x= (\tau,\theta,\phi_1,\phi_2)$ and $x'= (\tau',\theta',\phi_1',\phi_2')$.
We wish to compare the poles of the Fourier transform of this propagator to
the quasi-normal frequencies. We shall obtain agreement in a certain scaling
limit.

\subsection{The scaling limit}

As discussed above, the conformal group $SO(5,1)$ on the Euclidean space $\mathbb{R}\times S^3$
has a maximal subgroup $SO(1,1)\times SO(4)$. The representations of $SO(5,1)$
are labeled by the quantum numbers $(E,L)$ with respective Casimirs $E^2$
and $L(L+2)$ (eigenvalues of the respective Laplacians on $\mathbb{R}$ and $S^3$).
In the limit of large $E$ and $L$, we may approximate $L(L+2)
\approx (L+1)^2$. We shall consider the scaling limit that selects the
diagonal representation in the large $(E,L)$ limit, {\em i.e.,}
\be E \approx L+1\ee
On account of the quantization condition~(\ref{eqE}), we then obtain in this limit
$\pm iE \approx E+\Delta_++2n-3$, leading to
\be\label{eq79} iE \approx (-i\pm 1) (\half\Delta_+ + n - {\textstyle{\frac{3}{2}}})\ee
To implement this scaling limit, we may expand around the helix $\phi_1 = \tau [\mathrm{mod} 2\pi ]$, where $\tau$ is the coordinate in $\mathbb{R}$ and $\phi_1$ is one of the periodic coordinates in $S^3$;
\footnote{Of course, this is not a null geodesic. Since we are working in
Euclidean space, null geodesics do not exist.} the metric in
$\mathbb{R}\times S^3$ is
\be ds_{\mathbb{R}\times S^3}^2 = d\tau^2 +\cos^2\theta d\phi_1^2 + d\theta^2
+ \sin^2\theta d\phi_2^2\ee
and is easily obtained from the AdS$_5$ metric~(\ref{eq57}) by going to
the boundary ($u\to 1$).

Expanding
around the equator (small $\theta$) turns the sphere $S^3$ on which $\hat L^2$
(eq.~(\ref{eqL2})) acts into a cylinder $S^1\times \mathbb{R}^2$.
The Laplacian $\hat L^2$
(eq.~(\ref{eqL2})) then reads
\be \hat L^2 = - \frac{\partial^2}{\partial\phi_1^2} - \nabla_2^2\ee
where $\nabla_2^2$ is the Laplacian on $\mathbb{R}^2$. Assuming large angular
momentum in $S^1$ (quantum number $j_1$ in~(\ref{eqL2e})), the
momentum in $\mathbb{R}^2$ is negligible, and the Casimir is approximately
$\hat L^2 \approx j_1^2$.
Since we are expanding around the diagonal representation, in this limit we have
$j_1\approx E$.
If we identify the Euclidean time with
\be\label{eq74} t_E = \frac{\phi_1}{2\pi T}\ee
then we have periodicity under $t_E\to t_E + 1/T$ and the eigenfunction~(\ref{eqL2e}) of
$\hat L^2$ may be written as
\be e^{ij_1\phi_1} = e^{i\omega_E t_E} \ \ , \ \ \omega_E = 2\pi T j_1\ee
Since $j_1\approx E$, we conclude from~(\ref{eq79})
\be\label{eq76} i\omega_E \approx 2\pi T (-i\pm 1) (\half\Delta_+ + n - {\textstyle{\frac{3}{2}}})\ee
which is in agreement with the asymptotic form of quasi-normal frequencies~(\ref{eqA}).

Using $E\approx L+1\approx \frac{\omega_E}{2\pi T} \equiv -i\hat\omega$, the monodromies~(\ref{eqmoA})
become
\be\label{eqmoA1}\mathcal{M} (0) = \left( \ba{cc} e^{\pi (\hat\omega +i)} & \\ & e^{- \pi (\hat\omega -i)} \ea \right) \ , \ \mathcal{M} (1) = 
\left( \ba{cc} e^{i\pi \Delta_+} & \\ & e^{i\pi \Delta_-}\ea \right)  \ , \ \mathcal{M} (\infty) = \left( \ba{cc} e^{i\pi \hat\omega} & \\ & e^{-i\pi \hat\omega} \ea \right) \ee
They agree with the monodromies found in the high-energy limit for an AdS black hole~(eq.~(\ref{eqmono5})).
Thus, even though the two metrics~(\ref{sec3}) and (\ref{eq57}) are not locally equivalent, the $w$-surfaces
solving the corresponding wave equations in the limits considered are.

The above result may also be obtained by a direct
CFT calculation. To this end, we note that in the scaling limit in which the sphere $S^3$ labeled by $(\theta,\phi_1,\phi_2)$
turns into a cylinder $S^1\times \mathbb{R}^2$, where $\phi_1$ labels the circle $S^1$,
the invariant distance~(\ref{eq68}) becomes
\be \mathcal{P} \approx \cosh(\tau - \tau') - \cos (\phi_1 -\phi_1')\ee
which is of the same form as the two-dimensional expression~(\ref{eqPcal}).
The Green function is given by~(\ref{eq69}). Fourier transforming, we obtain
\be \tilde G \sim \int d\tau d\phi_1 e^{-i(E\tau +j_1\phi_1)}\ G(\tau,\phi_1;0,0)\ee
The expansion around the helix $\phi_1 = \tau$ selects the diagonal representation $E\approx j_1$. By a direct calculation, we conclude that the poles of $\tilde G$ in this approximation are located at
\be\label{eqonew} i\omega_E \approx 2\pi T (-i\pm 1) (\half\Delta_+ + n - 1)
\ \ , \ \ n = 1,2,\dots\ee
This is also easily derived by appealing to the discussion of the three-dimensional case and simply setting
\be \hat\omega = i\hat p = \frac{i\omega_E}{4\pi T}\ee
in eq.~(\ref{eqwme2}). Eq.~(\ref{eqonew}) is in agreement with the quasi-normal
frequencies~(\ref{eqA}) to leading order (both eqs.~(\ref{eqonew}) and (\ref{eqA}) are $o(m)$ expressions differing by a $o(1)$ term;
obtaining $o(1)$ agreement would entail a more careful derivation of the
CFT propagator in this scaling limit).

\section{Conclusions}\label{sec4}

We have discussed the connection between the quasi-normal modes of massive
scalar perturbations of AdS black holes in five dimensions and their
connection to the dual CFT in four dimensions.
On the gravity side, we obtained an analytic asymptotic expression for quasi-normal frequencies~(eq.~(\ref{eqA}))
valid at large $m$, where $m$ is the mass of the scalar field.
We did this by solving the wave equation, which reduced to a Heun equation~(\ref{w1}), perturbatively.
On the CFT side, we obtained an expression for the poles of the propagator by
expanding around the diagonal representation of the conformal group under its
maximal subgroup. Working in Euclidean space ($\mathbb{R}\times S^3$), this
can be seen as an expansion around a helix $\phi = \tau [ \mathrm{mod} 2\pi ]$,
where $\tau$ is the coordinate labeling $\mathbb{R}$ and $\phi$ is one of the periodic coordinates of the sphere $S^3$ (identified with Euclidean time; see eq.~(\ref{eq74})).
The asymptotic expression for the poles of the propagator thus derived~(eq.~(\ref{eq76});
see also~(\ref{eqonew})) was in agreement with the asymptotic form of the
quasi-normal frequencies derived on the gravity side~(eq.~(\ref{eqA})).

Our discussion generalized the three-dimensional case in which the correspondence is exact~\cite{bibq13}.
The geometric origin of this result in three dimensions is the local equivalence of the BTZ and AdS$_3$ metrics.
This guarantees the equivalence of the Riemann $w$-surfaces which solve the corresponding wave equations~\cite{bib26}.
In five dimensions, the black hole and AdS metrics are not equivalent.
The geometric origin of our result was, instead, the equivalence of the
corresponding $w$-surfaces in certain limits that we discussed.

Higher-order corrections may also be derived on both the gravity side and in CFT.
It would be interesting to see if agreement persists at higher orders.
It would also be interesting to extend the above results to more general black holes
in AdS$_d$ for $d\ne 3,5$.
The number of (unphysical) singularities increases, so the analysis
is expected to be considerably more involved.

%\section*{Acknowledgments}
%
%G.~S.~is supported by the US Department of Energy under grant
%DE-FG05-91ER40627.

\end{document}